\PassOptionsToPackage{colorlinks,citecolor=blue,linkcolor=blue,urlcolor=blue,bookmarks=false,hypertexnames=true}{hyperref}
\documentclass[floatfix,%
reprint,
superscriptaddress,
amsmath,
amssymb,
aps,
prx,
]{revtex4-2}

\usepackage{xcolor}
\usepackage[percent]{overpic} 
\usepackage{qcircuit}
\usepackage{graphicx}
\newcounter{mysubfig}[figure]

\usepackage{csquotes}

\usepackage{amsmath} 
\usepackage{graphicx}
\usepackage{placeins}
\usepackage[normalem]{ulem}
\usepackage{dsfont} 
\usepackage{upgreek}
\usepackage{booktabs}
\usepackage{tabularx}
\usepackage{tikz}
\usetikzlibrary{arrows.meta, positioning, fit}

\usepackage{bookmark}

\begin{document}

\title{Support Vector Machine with a Scalable Quantum Kernel}
\author{Anant Agnihotri}
\affiliation{Fraunhofer Institute for Applied Solid State Physics IAF, Tullastraße 72, 79108, Freiburg im Breisgau}
\author{Michael Krebsbach}
\affiliation{Fraunhofer Institute for Applied Solid State Physics IAF, Tullastraße 72, 79108, Freiburg im Breisgau}
\author{Florentin Reiter}
\affiliation{Fraunhofer Institute for Applied Solid State Physics IAF, Tullastraße 72, 79108, Freiburg im Breisgau}
\author{Thomas Wellens}
\affiliation{Fraunhofer Institute for Applied Solid State Physics IAF, Tullastraße 72, 79108, Freiburg im Breisgau}

\date{\today}

\begin{abstract}

Quantum support vector machines 
are classification algorithms that rely on quantum-generated kernels. The fidelity quantum kernel 
commonly used in 
quantum support vector machines 
suffers from exponential concentration as system size increases, preventing an efficient scaling beyond few-qubit systems. We introduce the 
Hamming quantum kernel,
a classical post-processing method that is based on the same measurement outcomes as the 
fidelity quantum kernel.
However, it avoids the exponential concentration problem by using the full measurement statistics 
rather than a single fidelity value. We evaluate the approach on both classical data (MNIST) and synthetic data generated from quantum circuits, using systems ranging from 2 to 27 qubits. Throughout the 
simulations,
the 
Hamming quantum kernel
outperforms the 
fidelity quantum kernel 
whenever 15 or more qubits are used. Furthermore, for synthetic quantum data, our method consistently outperforms the classical Gaussian kernel. This demonstrates that the 
Hamming quantum kernel
improves the expressivity and robustness at larger qubit scales without requiring any additional quantum ressources.
\end{abstract}

\maketitle

\section{Introduction}

Quantum machine learning is a fast-evolving area within quantum computing. Hybrid algorithms \cite{Bharti_2022}, where quantum processors handle only selected subroutines of a larger workflow have become particularly attractive. These methods offer a practical testbed for assessing quantum architectures on real-world tasks, leveraging quantum mechanical advantages while keeping overall resource requirements manageable. Even as research advances toward fault tolerant application scale (FASQ) quantum computers, hybrid approaches remain an effective means of exploring and validating quantum capabilities. 

Quantum support vector machines (QSVMs) \cite{PhysRevLett.113.130503} are notable examples of  hybrid supervised machine learning algorithms for data classification. Various benchmarking studies have been conducted to analyze their implementation and performance. 
Refs.~\cite{Havl_ek_2019} and \cite{schuld2019} proposed a quantum kernel estimator that computes 
kernel functions using quantum state space as a feature space, 
leveraging the exponentially large Hilbert space to potentially enhance a classical SVM for suitable datasets. A QSVM model to study the breast cancer dataset and evaluate its accuracy against a classical SVM was performed in \cite{bcancer}. The evaluation was performed on a dataset of 536 samples. Classical SVM achieved a test accuracy of 90\%, outperforming QSVM by 5 percentage points.  A systematic comparison of more than $20{,}000$ models to investigate the impact of various hyperparameters on kernel accuracy is carried out in \cite{Schnabel_2025}.  A key finding of the study emphasizes the critical role of hyperparameters, particularly regularization and bandwidth tuning, in ensuring the effective training of kernels.

Standard quantum kernel methods utilize the fidelity quantum kernel (FQK), which maps data points to quantum states known as feature vectors and computes their overlap in the Hilbert space. Early investigations 
of
fidelity quantum kernels demonstrate that they can, in principle, solve problems that cannot be efficiently addressed by classical computers \cite{Liu_2021}. However, this has only been demonstrated for highly specialized problems with structures that quantum computers can leverage, and it necessitates a FASQ quantum computer because it relies on Shor-type data encoding. General-purpose ansätze, on the other hand, suffer from the so-called exponential concentration \cite{thanasilp2024exponentialconcentrationquantumkernel}. Although they have shown promising results on a small scale \cite{muser2023provableadvantageskernelbasedquantum}, their reliance on calculating overlaps between different $n$-qubit quantum states hampers effective scaling to large $n$. Assuming that the feature vectors are randomly chosen $n$-qubit quantum states, their overlap scales like $\sim 2^{-n}$, which would require an exponential number of measurements to distinguish them from $0$. 

A solution to this problem was proposed by projected quantum kernels \cite{Huang_2021}, where the inner 
product
is calculated between low-dimensional approximations of the state. However, this approach limits the ability to capture high-order correlations between data points, making it potentially unsuitable for certain datasets. Ref. \cite{Slattery_2023} presents numerical results suggesting that fidelity quantum kernels (FQKs) do not show a clear quantum advantage when applied to classical data highlighting the impact of dataset for kernel methods. They show that tuning the kernel bandwidth can improve model performance and help with generalization by adjusting how sensitive the kernel is to differences in the data. However, this also makes the kernels easier to simulate classically, due to unfavorable geometric difference values \cite{Huang_2021}, which reduces the potential quantum advantage. 
\begin{figure*}[t]
    \centering
    \includegraphics[width=\textwidth]{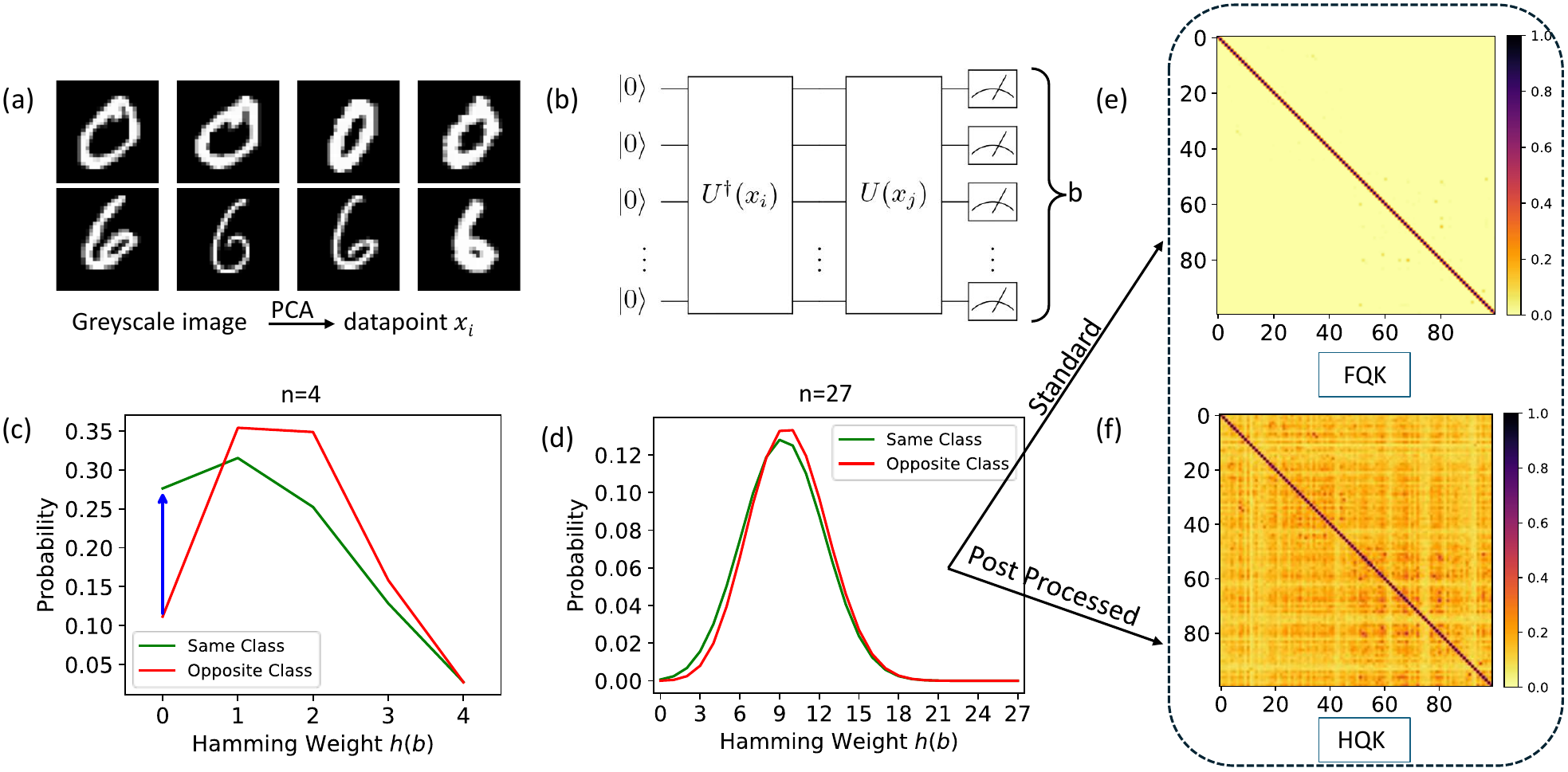}
    \caption{
    Illustration of the quantum kernel method presented in this work: (a) The datapoints $x_i$ are real vectors with $n=2$ to $n=27$ dimensions. We use different datasets including MNIST data downscaled by Principle Component Analysis, on the one hand and synthetic data originating from a quantum process, on the other hand. (b) Pairs of datapoints $x_i$ and $x_j$ are compared to one another using an $n$ qubit quantum circuit. The underlying assumption is that datapoints of the same class are more similar than datapoints of different classes, leading to a higher probability of measuring the all-zero bitstring. The FQK uses this probability as kernel entries, ignoring all other bitstrings. Instead, we use the additional heuristic that similar datapoints on average lead to bitstrings of low Hamming weight. For the example of classification between MNIST digits $0$ and $6$, (c) and (d) show the average Hamming distributions for $n=4$ and $n=27$ qubits for pairs of datapoints $x_i$ and $x_j$ belonging to the same (green) or to different classes (red). The blue arrow highlights the difference between in FQK kernel entries, corresponding to the information utilized by the SVM for classification. As the system size increases, the differences in the average FQK entry concentrate exponentially, leading to a sparse kernel matrix (e) that providing limited information for SVM optimization. However, the Hamming distributions are still distinctly shifted, allowing for an efficient classification. We use a classical post-processing method that leverages the full measurement statistics, yielding a more expressive and informative kernel (f) for SVM optimization.}
    \label{fig:your_label}
\end{figure*}

In this work, we propose a method called the Hamming quantum kernel (HQK), which computes 
kernel matrix elements
directly on the quantum computer while ensuring 
scalability. The key idea behind ensuring scalability is to utilize the complete measurement statistics,
rather than focusing on a single or a few specific bitstrings. 
For this purpose, 
classical post-processing leverages the distribution of the 
measured bitstrings
in a meaningful way, giving rise to distribution-informed post-processing functions. The recent work \cite{agliardi2024mitigatingexponentialconcentrationcovariant} concentrates on one such classical post-processing function, grouping the 
measured bitstrings 
by their Hamming weight. We similarly exploit the Hamming weight 
in
the measurement statistics and develop a post-processing function that 
increases
the 
similarity
between pairs of states. In particular, instead of focusing solely on the $|0\cdots0\rangle$ state, we apply our post-processing function to the complete measurement statistics grouped according to their Hamming weight, computing a weighted sum of the signal from each group to boost the similarity measure. Our method demonstrates improved accuracy on benchmark datasets compared to standard fidelity quantum 
kernels (FQKs) for feature vector dimensions greater than $15$, supporting the claim that it scales favourably. An overview of the method is illustrated in Fig.~\ref{fig:your_label}


We employ two distinct datasets to thoroughly evaluate our approach and conduct simulations involving up to $27$ qubits to examine the scalability of our method. The first dataset is the MNIST dataset (greyscale images of digits between $0$ and $9$) and the second dataset is a synthetically generated dataset from the outputs of a quantum circuit with different initial states. The MNIST dataset of handwritten digits is a well-studied dataset for classical machine learning algorithms. We use it here primarily to demonstrate the avoidance of exponential concentration in the proposed Hamming quantum kernel. Ref \cite{Belis_2024} demonstrates the superiority of quantum kernels over classical kernels for anomaly detection in high energy physics. Building on this success, we employ synthetically generated quantum data to show that quantum kernel methods can exceed classical approaches. Specifically, we use bitstrings from two quantum circuits, $V_0$ and $V_1$, as features for a quantum support vector machine to classify their source.

This work is structured as follows. Section \ref{background} introduces the theoretical foundations of quantum kernel methods, delving into the core principles of quantum kernel estimation, its operational mechanics, and the barriers encountered for scaling the estimation of the quantum kernel. Section \ref{method} outlines the methodology of our model, which integrates quantum kernel estimation with classical post-processing techniques, and describes the steps involved in implementing our approach. Section \ref{sec:experiments} details the datasets used for the comprehensive kernel analysis, including the preprocessing of classical data and the generation of quantum data along with the details about hyperparameter optimization and the test accuracy. In Section~\ref{Results}, we present a comparative analysis of kernel performance, evaluating the standard fidelity quantum kernel (FQK), the proposed modified Hamming quantum kernel (HQK), and the classical radial basis function (RBF) kernel (a Gaussian kernel) across multiple datasets.
By examining the performance differences between these kernels,  
we provide insights into the capabilities and limitations of different kernel methods in the context of quantum machine learning. Furthermore, we demonstrate that the proposed HQK achieves improved performance compared to the FQK on classical datasets, and outperforms both the RBF and FQK on quantum datasets.
Finally, in Section \ref{conc}, we summarize the insights gained from the proposed method and outline several open questions that merit further investigation.

\section{Background}
\label{background}

The support vector machine (SVM) is a supervised learning algorithm commonly employed for both classification and regression tasks. Its primary objective is to identify an optimal separating hyperplane (usually 
in
a high dimensional feature space) that effectively discriminates between distinct classes within a given dataset, particularly in the context of discrete classification problems. The implementation of SVM employs kernel functions that enable the calculation of the similarity metric between pairs of data points in the feature space without explicitly computing their coordinates. This approach, known as the kernel trick \cite{Hofmann_2008}, allows efficient handling of high-dimensional feature spaces.


\subsection{Kernels}

A kernel $\kappa$ is a positive 
semi-definite
function $\kappa: \chi \times \chi \rightarrow \mathbb{K}$ 
that takes two data points $x_i$ and $x_j$ from the input space $\chi$ and maps them onto $\kappa(x_i,x_j)\in \mathbb{K}$,
where $\mathbb{K}$ can be either $\mathbb{R}$ or $\mathbb{C}$. 
By definition, positive semi-definiteness of the kernel means that, for any set of data points $x_1,x_2,...,x_N\in\chi$, the corresponding $N\times N$-matrix 
$K(i,j)=\kappa(x_i,x_j)$
is positive semi-definite. 
According to Mercer’s theorem \cite{10.1007/11776420_14}, the positive semi-definiteness of kernels is equivalent to the existence of a feature map $\phi : \chi \rightarrow \mathcal{H}$, where $\mathcal{H}$ is a possibly high (or even infinite)-dimensional feature space. The relationship between the kernel function and the feature map is expressed as $\kappa(x, x^{\prime}) = \langle \phi(x), \phi(x^{\prime}) \rangle$, where $\langle \cdot , \cdot \rangle$ denotes the inner product in $\mathcal{H}$.

\subsubsection{Classical Gaussian RBF kernel}

In classical kernel methods, mappings from the input space to a feature space are implicitly defined by positive-definite kernel functions, such as the linear, polynomial, and radial basis function (RBF) kernels \cite{Rasmussen2006Gaussian}. Among these, the RBF kernel is particularly common in SVM classification. For two data points $x_i, x_j$ in the input space, it is defined as

\begin{equation}
    \kappa(x_i, x_j) = \exp\left(-\gamma\lVert x_i - x_j \rVert^2\right),
    \label{eq:rbf}
\end{equation}
where $\lVert x_i - x_j \rVert^2$ denotes the squared Euclidean distance and $\gamma > 0$ is a bandwidth parameter. The kernel value $\kappa(x_i, x_j)$ decreases monotonically with distance and lies in the interval $[0,1]$, which allows it to be naturally interpreted as a similarity measure. The parameter $\gamma$ controls the locality of this similarity, with larger values yielding more localized similarities and smaller values producing smoother, more global similarities.


\subsubsection{Standard fidelity quantum kernel}

In the quantum setting, feature maps are implemented by parameterized quantum circuits $U(x)$ that embed input data $x$ into a Hilbert space: 
\begin{equation}
|\psi(x)\rangle = U(x)|0\rangle^{\otimes n} ~,
\end{equation}
where $|0\rangle^{\otimes n}$ denotes the  all-zero computational basis state.
Although various circuit architectures can serve as feature maps, highly entangling circuits are often preferred because they are more difficult to simulate classically and may capture correlations among data points that are inaccessible to classical kernels. A widely used option in quantum machine learning is the ZZ kernel \cite{Havl_ek_2019},

\begin{equation}
\begin{split}
U{(x_i)} = \bigg[\exp \Biggl( &i \sum_{k=0}^{n-1} \phi_k(x_{i}^k) Z_k \\
&+ i \sum_{k=0}^{n-2} \phi_{k,k+1}(x_{i}^k, x_{i}^{k+1}) Z_k Z_{k+1} \Biggr) H^{\otimes n} \bigg]^{m},
\end{split}
\label{eq:Uxi}
\end{equation}
where $H^{\otimes n}$ denotes the Hadamard transform, $x_{i}^k, x_{i}^{k+1}$ are the elements of the vector $x_i$, $Z_k$ is the Pauli-Z operator on qubit $k$, and the mapping functions are $\phi_k(x_{i}^k) = x_{i}^k$ and $\phi_{k,k+1}(x_{i}^k, x_{i}^{k+1}) = (\pi - x_{i}^k)(\pi - x_{i}^{k+1})$. Every datapoint $x_i$ is scaled between $0$ and $2\pi$ before being encoded onto the feature map. The parameter $m$ denotes the number of feature map repetitions, which corresponds to the number of times the input data is re-uploaded into the model. Repeated data re-uploading into a quantum model has been shown to enhance its expressivity 
\cite{aminpour2024strategicdatareuploadspathway}. In this work, the feature map is applied three times($\text{m}=3$), re-uploading the data through three successive layers of the circuit.

The FQK is then given by the overlap between the respective states,
i.e., $\kappa_{\rm FQK}(x_i,x_j)=|\langle\psi(x_i)|\psi(x_j)\rangle|^2$,
giving rise to the kernel matrix
\begin{equation}
    \kappa_{\rm FQK}(x_i,x_j) = \left|\langle 0|^{\otimes n} U^{\dagger}(x_i) U(x_j) |0\rangle^{\otimes n}\right|^2.
\label{eq:FQK}
\end{equation}
Effectively, the overlap between two states is computed by preparing $|\psi(x_i,x_j)\rangle = U^{\dagger}(x_i) U(x_j) |0\rangle^{\otimes n}$ and measuring the overlap with 
the all-zero state $\langle 0|^{\otimes n}$.
Alternatively, a Hadamard swap test \cite{barenco1996stabilisationquantumcomputationssymmetrisation} can also be used to measure the overlap at the cost of more qubits and long range entangling gates. 

A practical concern with quantum kernels is ensuring that they are positive semidefinite (PSD), which is necessary for valid kernel-based learning. In the presence of sampling noise due to a finite number of shots, however, the estimated kernel matrix may be non-symmetric (and hence also not positive semi-definite). 
While a symmetric kernel $K^{(S)}$ can be enforced by defining each entry as 
$K^{(S)} (i, j) = [K(i, j) + K(j, i)]/2$
this would require evaluating approximately twice the number of circuits. To mitigate this computational overhead, we construct only the upper triangular part of the kernel matrix and define the lower triangle by symmetry, i.e., 
\( K(i, j) = K(j, i) \). 
Although this enforces symmetry, it does not guarantee positive semidefiniteness. PSD can be achieved for any symmetric or Hermitian matrix by projecting it onto the nearest PSD matrix. The projection is realized by eigendecomposing the matrix and replacing its negative eigenvalues by $0$. The same procedure to guarantee positive semidefinitness of the quantum kernel for system size less than $10$ will also be used later for our Hamming quantum kernel (See Sec. \ref{method}) as the systems exceeding $10$ qubits satisfy the PSD condition without the need of this procedure. 

The performance of a kernel method strongly depends on the choice of kernel, and no single kernel is universally optimal across all problem settings. One common approach to selecting an appropriate kernel is \textit{kernel alignment} \cite{NIPS2001_1f71e393}, which quantifies the similarity between a given kernel and a target kernel derived from the ground truth labels. Kernel alignment, denoted \(A(K_t, K)\), measures the agreement between the two kernels, where \(K_t\) is computed using only the training subset due to the availability of label information. A high alignment score indicates that the kernel is well-suited to the problem. In practice, kernels are often parameterized as $\left|\langle 0|^{\otimes n}U^\dagger(x_i,\theta)\,U(x_j,\theta)|0\rangle^{\otimes n}\right|^2$, with parameters \(\theta\) optimized to maximize alignment. In the ZZ kernel, kernel alignment can be performed by parameterizing the mapping function $\phi$ in Eq.~[\ref{eq:Uxi}] after the input data is encoded. While this procedure can yield effective kernels, it requires optimizing kernel parameters in an iterative setting, which significantly increases the computational cost.  Furthermore, it is not guaranteed to find good kernels due to problems like barren plateaus \cite{Larocca2025}, local minima, or the exponential concentration of kernel entries described in Sec. \ref{barriers}. In this work, we will therefore not apply kernel alignment and use the ZZ-kernel as introduced in Eq.~[\ref{eq:Uxi}].

\subsection{Barriers to Scalability in FQK}
\label{barriers}

FQK for QSVM suffers from a major downside: kernel entries exponentially concentrate with increasing system size \cite{thanasilp2024exponentialconcentrationquantumkernel}, driving the kernel 
matrix toward the identity and yielding near-random performance. As the problem size increases, the differences between kernel values become progressively smaller, making it more challenging to reliably distinguish between them. Consequently, a larger number of measurement shots is required to resolve these differences. Under a polynomial shot budget, this can result in a model that is largely insensitive to the input data and exhibits poor generalization performance. It is defined as \cite{thanasilp2024exponentialconcentrationquantumkernel},
\begin{equation}
    Pr_\alpha[|X(\alpha) - \mu|\geq \delta] \leq \frac{\beta}{\delta^2}, \beta \in \mathcal{O}(1 / b^n).
\end{equation}
with $b>1$. Here, $X(\alpha)$ denotes a quantity that depends on a set of variables $\alpha$ and can be evaluated on a quantum computer as the expectation value of a given observable. The quantity $X(\alpha)$ is said to exhibit exponential concentration in the number of qubits $n$ if the probability that it deviates from $\mu$ by a small amount $\delta$ is exponentially small. In the quantum kernel case, $X(\alpha)$ is equivalent to $\kappa(x_i, x_j)$, where $\alpha$ is a set of data points $x_i$ and $x_j$ (with $x_i\neq x_j$) embeded onto the quantum kernel. Probabilistic exponential concentration then implies that the probability of kernel matrix entries deviating from a fixed value $\mu$ that is independent of the input data is exponentially small, which in practice means that all non-diagonal entries converge to $0$ exponentially fast with high probability as the number of qubits $n$ increases, 




In section \ref{method}, we present an alternative method that mitigates the effect of exponential concentration on kernel entries and at the same time does not require hybrid optimization loops for the calculation of the final kernel.

\section{Method}
\label{method}

The fidelity-based quantum kernel defined in Eq.~(\ref{eq:FQK}) faces increasing difficulty in resolving differences between input states as the number of qubits grows, posing a practical challenge for quantum machine learning models. We require an approach to increase the overlap signal among pairs of data points in a meaningful way. The new approach should leverage the probability distribution derived from the measurement statistics of the encoded input data overlap to develop 
distribution-informed post-processing methods. We propose a heuristic approach, as one of the many possible distribution informed post processing functions, that leverages the full overlap statistics between pairs of quantum states instead of the frequency of measuring the all zero $|00.....0\rangle$ bitstring. Specifically, we group measurement outcomes (bitstrings) by their Hamming weight and assign a weight to each group, modulating its influence on the kernel value:


\begin{equation}
\kappa_{HQK}(x_i,x_j) = \sum_b |\langle b|U^{\dagger}(x_i)U(x_j)|0\rangle^{\otimes n}|^2 e^{-\lambda h(b)} ~.
\label{eq:HQK}
\end{equation}
Here, $h(b)$ is the Hamming weight of the bitstring $b$, and $\lambda$ is a hyperparameter. In effect, $\lambda$ controls how much of the full measurement statistics contribute to the similarity measure between data points. This method ensures that, as the system size increases,  the similarity metric between 
different
data points 
will not decrease to zero.  
The kernel matrix \( \kappa_{\rm HQK} \) will have entries all equal to $1$ when \( \lambda = 0 \) and asymptotically converges to the fidelity-based quantum kernel $K_{\rm FQK}$, see Eq.~(\ref{eq:FQK}), as \( \lambda \rightarrow \infty \). In the following, we treat \( \lambda \) as a tunable hyperparameter and optimize it in post processing to obtain the best-performing kernel. 



Finally, we note that the use of Hamming weights for error mitigation in quantum kernels was first introduced in \cite{agliardi2024mitigatingexponentialconcentrationcovariant}. They investigate another way of utilizing the complete measurement statistics based on hamming weight. Our approach thus differs in how the Hamming weight information is incorporated into the post-processing step, offering a distinct strategy for adjusting kernel values based on the overlap statistics. By relying on all measured bitstrings, discounted by an appropriate weight, instead of only few, selected bitstrings close to the all-zero bitstring, our approach can seamlessly deal with different problems and system sizes without requiring to adjust the kernel.

\section{Simulations}
\label{sec:experiments}

In the following, we will compare the performance of the three kernels explained above, i.e., the classical RBF Gaussian kernel, 
the standard quantum fidelity kernel (FQK) and the
Hamming quantum kernel (HQK), 
for different classification tasks based on corresponding datasets of varying dimensionality. After explaining the datasets and corresponding classification tasks in Sec.~\ref{sec:datasets},
we describe our procedure to find a suitable value of the hyperparameter $\lambda$. The test accuracy as our figure-of-merit for comparing the different kernels is finally introduced in Sec.~\ref{sec:test_accuracy}.  

\subsection{Datasets and classification tasks}
\label{sec:datasets}

The datasets used to assess the quality of the model are broadly divided into two categories, namely classical and quantum datasets. 

\subsubsection{Classical Dataset}

The classical datasets are based on the 
MNIST dataset \cite{6296535}, which consists of greyscale images of handwritten digits from $0$ to $9$.
For empirical evaluation, the MNIST dataset is partitioned into three binary classification tasks of increasing difficulty:
\begin{itemize}
    \item \textbf{Task 1:} Distinguish between the digits \textit{0} and \textit{1}.
    \item \textbf{Task 2:} Distinguish between the digits \textit{0} and \textit{6}.
    \item \textbf{Task 3:} Classify each digit as \textit{odd} or \textit{even}.
\end{itemize}
This progressive setup enables a systematic assessment of model performance as the classification challenge becomes more complex.

In each experimental simulation run, we take 100 labeled images as training dataset, 50 images of \textit{0} and 50 images of \textit{1} for task 1, 50 images of \textit{0} and 50 images of \textit{6} for task 2, and 50 randomly chosen images of digits chosen from the sets, $\{0,2,4,6,8\}$ and $\{1,3,5,7,9\}$ labeled as \textit{odd} or \textit{even} for task 3. 
Each image consists of $28\times28$ pixels, amounting to a 784-dimensional data point. We apply 
principal component analysis 
(PCA) \cite{MACKIEWICZ1993303} to reduce the dimensionality $n$ of the data points $x$ to a range between $2$ and $27$.

\begin{figure*}
    \begin{center}
    \begin{minipage}{0.5\textwidth}
        \centering
        \raisebox{-0.5\height}{(a)\label{fig:qa}}%
        \hspace{1.3em}%
        \raisebox{-0.5\height}{%
        \resizebox{0.9\textwidth}{!}{%
        \Qcircuit @C=0.7em @R=1em {
          \lstick{|0\rangle} \gategroup{1}{2}{6}{2}{1em}{--}
            & \gate{X^\alpha} & \qw
            & \gate{H} \gategroup{1}{4}{6}{14}{.7em}{--}
            & \ctrl{1} & \qw            & \ctrl{1} & \qw      & \qw            & \qw    & \qw                    & \qw            & \qw                    & \gate{R_x(\theta)} & \qw & \meter \gategroup{1}{16}{6}{16}{0.9em}{--} \\
          \lstick{|0\rangle}
            & \gate{X^\alpha} & \qw
            & \gate{H}
            & \targ    & \gate{P(\phi)} & \targ    & \ctrl{1} & \qw            & \ctrl{1} & \qw                  & \qw            & \qw                    & \gate{R_x(\theta)} & \qw & \meter \\
          \lstick{|0\rangle}
            & \gate{X^\alpha} & \qw
            & \gate{H}
            & \qw      & \qw            & \qw      & \targ    & \gate{P(\phi)} & \targ  & \qw                    & \qw            & \qw                    & \gate{R_x(\theta)} & \qw & \meter \\
          & \vdots & & \vdots & & \vdots & & & \vdots & & & \vdots & & \vdots & & \vdots \\
          & & & & & & & & & & *{\bullet}\ar@{-}[1,0] & & *{\bullet}\ar@{-}[1,0] & & & \\
          \lstick{|0\rangle}
            & \gate{X^\alpha} &\qw
            & \gate{H}
            & \qw      & \qw            & \qw      & \qw      & \qw            & \qw    & \targ                  & \gate{P(\phi)} & \targ                  & \gate{R_x(\theta)} & \qw & \meter \\
          & & & & & & & & & & & & & & & \\
          & \push{\text{(I)}} & & & & & & & \push{\text{(II)}} & & & & & & & \push{\text{(III)}} \\
        }
        }}
    \end{minipage}%
    \hfill
    \begin{minipage}{0.45\textwidth}
        \centering
        \raisebox{7\height}{(b)\label{fig:qb}}%
        \hspace{0.3em}%
        \raisebox{-0.5\height}{\includegraphics[width=0.9\textwidth]{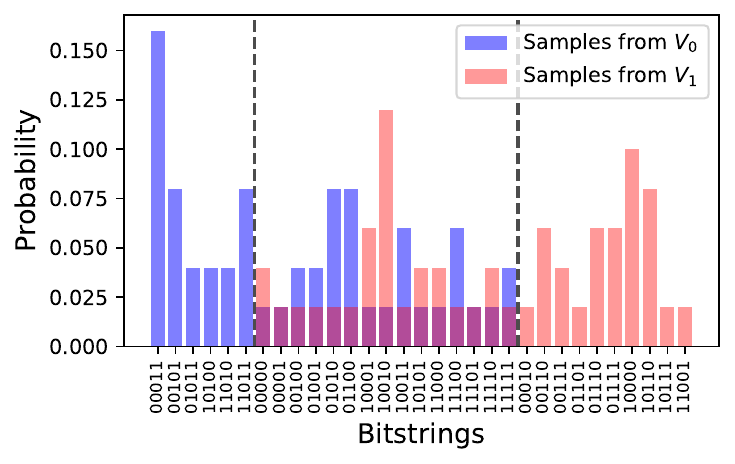}}
    \end{minipage}
    \end{center}
    \caption{
    (a) Quantum circuits $V_\alpha$ ($\alpha =0$ or $1$) used to generate quantum datasets. 
    (I) Depending on the binary label $\alpha$, the initial state is either $|0\rangle^{\otimes n_q}$ (for $\alpha=0$) or $|1\rangle^{\otimes n_q}$ (for $\alpha=1$), $n_q$ is the number of qubits in the circuit.
    (II) A sequence of $ZZ$-rotations, i.e., $\exp(-i \frac{\phi}{2} Z\otimes Z)$ realized by two CNOT gates with phase gate $P(\phi)$ applied on the target qubit in between, is applied between qubit pairs $i,i+1$ ($i=1,\dots,n-1$). Then, a layer of $R_x(\theta)$-rotations is applied to all qubits. 
    The values of the parameters are: $\phi = 0.8\pi$, $\theta= 0.64\pi$. 
    (III) The final measurement yields a bitstring $x$ (quantum data point). The classification task consists of finding out whether a given bitstring $x$ has been generated using $V_0$ or $V_1$. 
    (b) Probability distributions of bitstrings sampled from circuits $V_0$ and $V_1$ for $n=5$ qubits. Blue bars represent bitstrings generated by $V_0$, while pink bars represent  those generated by $V_1$. The overlapping region highlights bitstrings common  to both circuits. The classification task of the SVM is to distinguish between the two  distributions, i.e., to identify the circuit $V_0$ or $V_1$ from which a given bitstring originates.
    }
    \label{fig:qdata}
\end{figure*}
\subsubsection{Quantum Dataset}
The second category of dataset used in this assessment is a synthetically generated quantum dataset (whereby we mean a classical dataset generated by a quantum device). This dataset consists of binary bitstrings $x$ produced as measurement outcomes by one of two quantum circuits $V_\alpha$, where $\alpha=0$ or $1$,
see Fig.~\ref{fig:qdata}. The classification task is to determine whether a given bitstring $x$ has been generated using $V_0$ or $V_1$. For each experimental run, we generate a training dataset with 100 quantum data points (50 bitstrings from circuit $V_0$ and 50 from $V_1$).
As this dataset is generated from a quantum circuit, scaling is performed by increasing the number $n$ of qubits in the data generation circuit
from $4$ to $27$. 

Beyond the diversity of the dataset, a further motivation for its use is the expectation that quantum data will yield superior performance on quantum models, given that such data is inherently more native to the quantum framework itself.
Importantly, the circuits $V_\alpha$ used to generate the quantum data are different from the circuit $U(x)$ used as feature map. Similar to the latter, the circuits $V_\alpha$ consist of $R_{ZZ}(\phi)$-gates as engangling gates. However, they additionally contain $R_X(\theta)$ rotation gates, with the rotation angles $\phi$ and $\theta$ for both sets of gates kept fixed. In contrast, the rotation angles $\phi_{kl}$ in $U(x)$, see Eq.~(\ref{eq:Uxi}), depend on the given data point $x$, which, in turn, is obtained as the measurement outcome of circuit $V_\alpha$.
Additionally, the quantum data generating circuit contains only a single layer of $RZZ$ gates, see Fig.~{\ref{fig:qdata}(a)}, part (II), whereas the feature map circuit $U(x)$ uses $\text{m}=3$ repetitions.
The fact that the circuits used for generating the quantum data points, on the one hand, and mapping them to the feature space, on the other hand, are different and not related to each other in an obvious way
avoids an unfair bias of the quantum 
kernels
towards this dataset. 


\subsection{Hyperparameter optimization}
\label{sec:lambda_opt}

The hyperparameter $\lambda$ determines the proportion of measurement statistics used in computing the Hamming quantum kernel element $\kappa_{\rm HQK}(x_i,x_j)$, see Eq.~(\ref{eq:HQK}). Expectation values utilized for determining  quantum kernel elements are always calculated over $1024$ shots (independently of the number of qubits).  
Hyperparameter optimization must not be performed on the test dataset, as this could potentially lead to overfitting: the model effectively \enquote{memorizes} the test data rather than genuinely learning from it. Instead, an independent subset of the data is required for hyperparameter selection. In this work, we employ one of the standard approaches for this purpose, namely $k$-fold cross-validation \cite{kfold}.
 


The training dataset, comprising 100 samples, is partitioned into five folds for cross-validation. In each fold, 80 samples are used for training and the remaining 20 samples for cross-validation. The validation subset is selected sequentially from the dataset, resulting in distinct training and validation splits for each fold. 

For every fold, the model is trained on the corresponding 80 training samples and evaluated on the 20-sample validation set. The hyperparameter $\lambda$ is tuned over the range $10^{-2}$ to $10^{2}$ on a logarithmic scale, in order to cover the entire range between the limiting cases $\lambda\to 0$, on the one hand, and $\lambda\to\infty$, on the other hand. 
Each fold is repeated five times to account for variability, and the cross-validation accuracy,
defined as the fraction of correctly classified samples of the validation set, 
is averaged across runs.

This procedure is 
repeated for different numbers of qubits (from $n=2$ to $27$). The hyperparameter yielding the highest overall average cross-validation accuracy is selected as optimal and subsequently used for evaluation on the held-out test set.

We perform this hyperparameter optimization for all experimental simulation runs conducted below. According to the 
results shown in Sec.~\ref{Results},
choices for $\lambda$ between $0.05$ and $0.3$, independently of the number of qubits, seem to reliably lead to good classification performances.

The RBF kernel is used as a classical baseline to assess the performance of our model. The RBF kernel is the standard kernel provided by \textit{scikit-learn}. The value of $\gamma$ (see Eq.~[\ref{eq:rbf}]) used in the kernel is the default value provided in \textit{scikit-learn} called \textit{scale}, defined as 
\begin{equation}
    \gamma = \frac{1}{n ~ \text{Var}(X)},
\end{equation}
where $n$ denotes the number of features and $\text{Var}(X)$ is the 
variance of the dataset defined as
\begin{equation}
    \text{Var}(X) = \frac{1}{n~m} \sum_{i=1}^{n} \sum_{k=1}^{m} (x_i^k - \bar{x})^2
\end{equation}
Here, $k$ goes from $1$ to the number of features in a single datapoint and $i$ goes from $1$ to the number of samples in the dataset. 
Datasets with large variances would have a smaller $\gamma$ leading to the RBF kernel having a wider spread and a much smoother boundary. Similarly, datasets with small variances would have a larger $\gamma$ leading to the RBF kernel having a narrow spread and a much tighter boundary. Additionally, an independent hyperparameter cross validation was also run on other $\gamma$ values ranging from $0.01$ to $100$ along with \textit{scale} and \textit{auto}. Across all feature dimensions and datasets, we found that the default value (\textit{scale}) given above indeed displays the best performance. It is therefore used in the following for all our classical RBF kernel calculations.


\subsection{Test accuracy}
\label{sec:test_accuracy}

For the tasks with classical (MNIST) data, we use a subset of $100$ images for training and $99$
randomly chosen 
images for testing. 
Principal component analysis (PCA) \cite{MACKIEWICZ1993303} is applied to reduce the dimensionality of the data from $784$ to a range of $2$ to $27$ dimensions, enabling a systematic scaling analysis across all three kernel methods: FQK, HQK, and RBF. The number $n$ of qubits  required to encode the data in the quantum kernels is equal to the dimensionality of each sample. For the quantum dataset, the dimensionality is directly varied by increasing the number of qubits in the generating circuits $V_\alpha$ (from $n=4$ to $27$).

Test accuracy is defined as the fraction of previously unseen data correctly classified by the trained model. To ensure robustness, test accuracy is evaluated over $5$ independent runs, with each run employing a different randomly selected subset of samples for the training and the test set. The reported test accuracy represents the mean value across all runs.

\section{Results and Discussion}
\label{Results}


For each classification task,
we first discuss the results of the hyperparameter optimization, 
and then analyze and compare the performance of the three kernels, i.e. RBF, FQK and the HQK, with increasing number $n$ of qubits (or, equivalently, data dimensionality).

\subsection{Kernel Quality for MNIST data}

We begin with the classical MNIST dataset, where, as explained in Sec.~\ref{sec:datasets}, we consider three classification tasks of increasing difficulty: distinguishing between 0/1, 0/6 and odd/even, respectively.

  
  

\subsubsection{Kernel Quality for 0/1}

\begin{figure}
    \centering

    \refstepcounter{mysubfig}\label{fig:plots01-a}%
    \begin{overpic}[width=\linewidth]{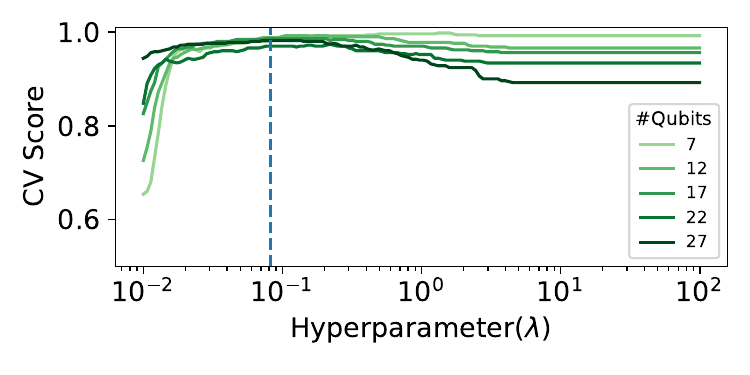}
        \put(-3,45){\scriptsize(a)}
    \end{overpic}

    \vspace{0.5em}

    \refstepcounter{mysubfig}\label{fig:plots01-b}%
    \begin{overpic}[width=\linewidth]{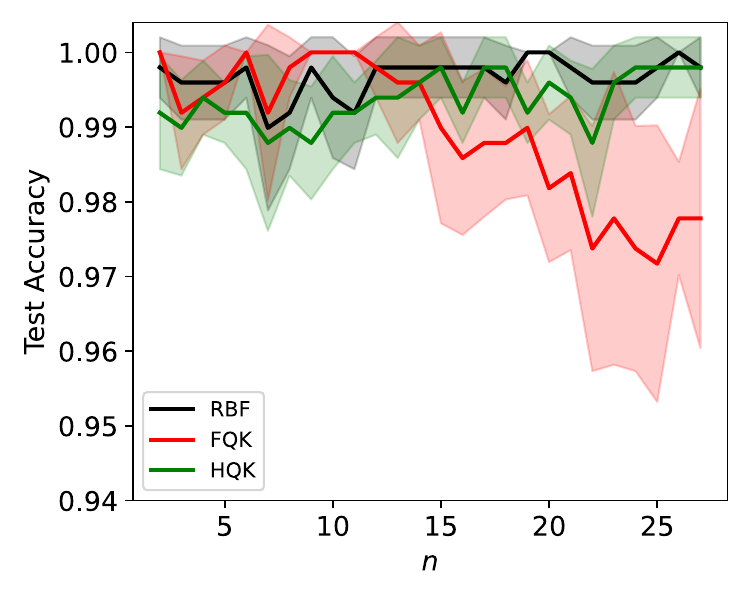}
        \put(-3,75){\scriptsize(b)}
    \end{overpic}

    \caption{Comparison of test scores 
    for classification of MNIST digits $0$ and $1$. 
    (a) HQK hyperparameter optimization: Cross-validation scores 
    as a function of $\lambda$ for different numbers of qubits (see inset). 
    The value of $\lambda$ selected for the test set evaluation is $0.08$ (vertical dashed line). (b) Test accuracy for the different kernels 
    (RBF: classical Gaussian kernel, FQK: standard fidelity quantum kernel and HQK: Hamming quantum kernel, see inset) 
    as a function of data dimensionality $n$ (or, equivalently, number of qubits used for the quantum kernels). 
    Mean values (solid lines) and standard deviations (shaded regions) for five independent runs are shown.
    For all kernels, two principal components ($n=2$) suffice for near-perfect classification of $0$ and $1$. This underlines the simplicity of this task, 
    as the digit $1$ is among the easiest to classify. As expected, FQK accuracy drops for larger $n$, whereas HQK remains stable and approaches RBF performance.}
    \label{fig:plots01}
\end{figure}

 First, we estimate the optimum value of
 the HQK hyperparameter 
 $\lambda$.
Fig.~\ref{fig:plots01-a} illustrates the cross-validation scores for qubit numbers $n=7, 12, 17, 22$ and $27$ as a function of $\lambda$. 
As mentioned in Sec.~\ref{method},
the HQK 
converges
to the standard FQK
for $\lambda \rightarrow \infty$, 
thus encountering the problem of exponential concentration with increasing $n$. 
Indeed, in the regime of large $\lambda$, the  cross-validation accuracy (CV score) decreases with an increasing number of qubits. This behaviour is also consistent with the use of PCA for dimensionality reduction. PCA projects data along directions of maximum variance, ordered from highest to lowest. The data encoded by $7$ qubits already achieves a CV score close to 1, i.e., it captures enough relevant information from the original dataset. Adding more dimensions contributes more noise than useful information, resulting in lower accuracy compounded by exponential concentration.

Conversely, small values of $\lambda$ cause each kernel entry to approach $1 - \epsilon$, 
with entries differing only at very high decimal precision. Despite their near-uniformity, 
these marginal differences($\epsilon$) in kernel values are sufficient for the SVM to achieve a 
cross-validation accuracy exceeding that of random guessing($0.5$).
However, an intermediate regime exists where an increase in cross-validation accuracy is observed 
for
all qubit.
The value of $\lambda$ selected for the test set evaluation is $0.08$ but any $\lambda$ within the range $0.05$--$0.5$ would yield similar results. 



After determining the optimal hyperparameter, a comparative analysis of test accuracy is performed on the dataset. Fig.~\ref{fig:plots01-b} shows the kernel accuracy for the test dataset.

For small $n$, the accuracy of all three methods is close to $1$. Hence, already two principal components ($n=2$) suffice for near-perfect classification of $0$ and $1$. 
For larger $n$, a decrease of the FQK test accuracy is observed, as a consequence of exponential concentration as discussed above. In contrast, HQK remains stable and achieves a similar accuracy as the classical RBF kernel. 

This is the first sign that 
the HQK
does not suffer from the limitations of the FQK. However, the present dataset is not sufficiently complex to evaluate the robustness of our approach. To overcome this limitation, we subsequently investigate more challenging datasets to more comprehensively evaluate its performance.

\begin{figure}
    \centering

    \refstepcounter{mysubfig}\label{fig:plots06-a}%
    \begin{overpic}[width=\linewidth]{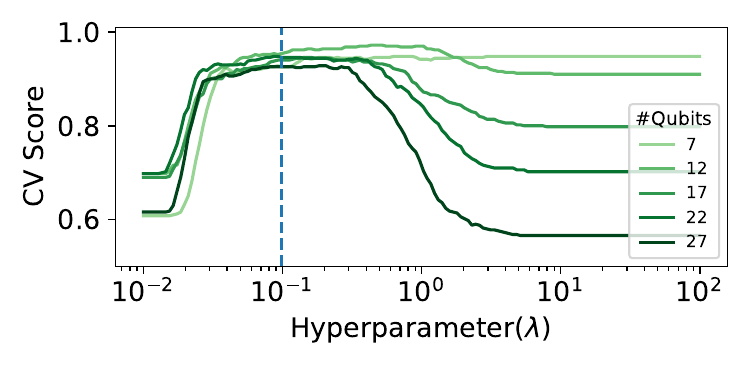}
        \put(-3,45){\scriptsize(a)}
    \end{overpic}

    \vspace{0.5em}

    \refstepcounter{mysubfig}\label{fig:plots06-b}%
    \begin{overpic}[width=\linewidth]{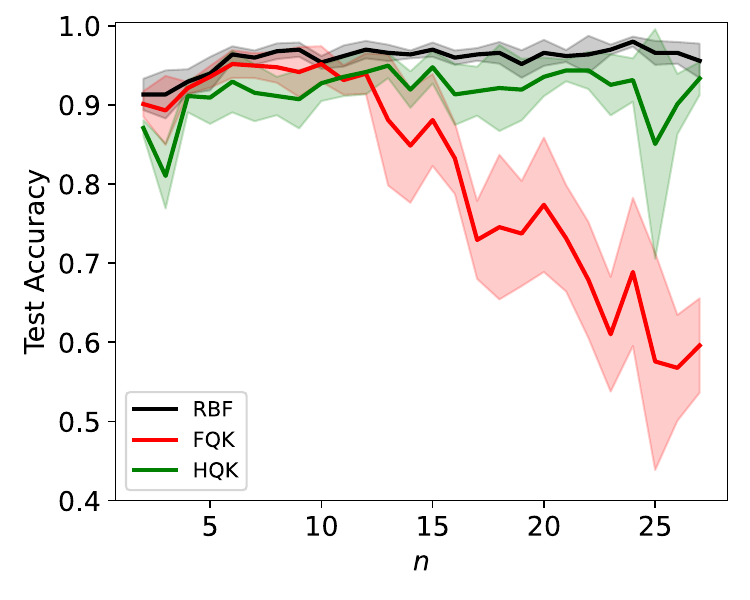}
        \put(-3,75){\scriptsize(b)}
    \end{overpic}

    \caption{
    Same as Fig.~\ref{fig:plots01},
    for classification of MNIST digits $0$ and $6$. 
    (a) 
    The optimized value of the HQK hyperparameter $\lambda$ for this dataset is $0.1$ (vertical dashed line). (b)
    Maximum test accuracy is achieved at eight principal components ($n=8$), enabling near-perfect classification of $0$ and $6$. 
    Beyond $n=8$, additional dimensions introduce more noise than relevant information. FQK accuracy drops for larger $n$, whereas HQK remains relatively stable and almost reaches 
    RBF performance, demonstrating higher robustness to noisy data than FQK.}
    \label{fig:plots06}
\end{figure}

\subsubsection{Kernel Quality for 0/6}

A similar binary classification is performed on a dataset of intermediate difficulty. This dataset also consists of $100$ training and $99$ test samples, comprising digits $0$ and $6$. We perform the same hyperparameter optimization to extract the optimal value of $\lambda$ that maximizes the cross-validation score, see Fig.~\ref{fig:plots06-a}.

The cross-validation accuracy 
exhibits qualitatively similar behavior to the $0/1$ dataset. There exists a regime between the extremes of FQK ($\lambda\to\infty$) and a kernel with all entries equal to $1$ ($\lambda\to 0$), where the cross-validation accuracy shows a peak. The accuracy begins to rise at $\lambda \sim 10^{-1.8}$ and reaches its maximum in the range $\lambda = 0.03$ to $\lambda = 0.3$. The optimized value of $\lambda$ chosen for this dataset is $0.1$.
As compared to the $0/1$ dataset, the drop in cross-validation accuracy for large $\lambda$ is steeper, especially for higher numbers of qubits.
This can be attributed to three main factors: the nature of the dataset, the ansatz used, and exponential concentration. More noise is introduced as the dimensionality of the dataset increases.

Fig.~\ref{fig:plots06-b} shows the test accuracies for the different kernels for the 0 vs.\ 6 classification task. The classical RBF kernel 
exhibits
the best classification accuracy for all 
values of $n$.
However, for 
low $n$
the quantum models achieve similar accuracies. For $n\geq 15$, the test accuracy of the FQK drops sharply, approaching a random guesser (corresponding to test accuracy $1/2$) for $n\geq 25$ qubits. This is a clear indication for exponential concentration in the FQK. The classification between digits $0$ and $6$ is inherently more challenging than between digits $0$ and $1$, and $1024$ measurement shots are insufficient to reliably estimate the overlaps between quantum states encoded with more than $n = 15$ qubits.

The HQK is able to avoid this drastic drop in accuracy. The weighting of bitstrings by Hamming weight helps avoiding the exponential concentration problem and enables efficient learning of this dataset with quantum kernels on a scale.

\subsubsection{Kernel Quality for odd/even}

The behavior of the HQK kernel for the binary classification of $0/6$ for large number of qubits exhibits a better performance than the standard FQK. We would now like to analyze the behavior of all three kernels on the hardest classical dataset among the three. 

\begin{figure}[htbp]
    \centering

    \refstepcounter{mysubfig}\label{fig:plotsoe-a}%
    \begin{overpic}[width=\linewidth]{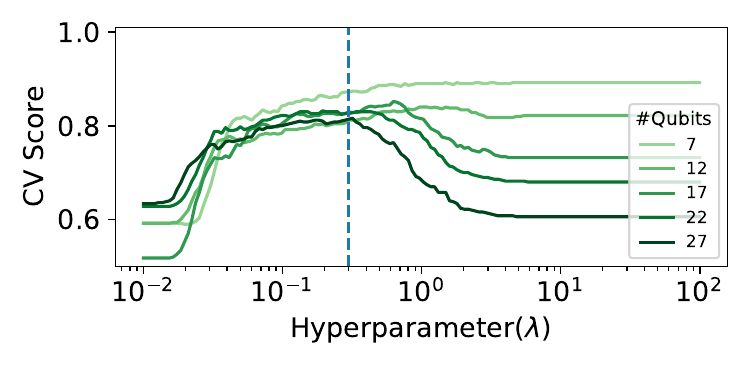}
        \put(-3,45){\scriptsize(a)}
    \end{overpic}

    \vspace{0.5em}

    \refstepcounter{mysubfig}\label{fig:plotsoe-b}%
    \begin{overpic}[width=\linewidth]{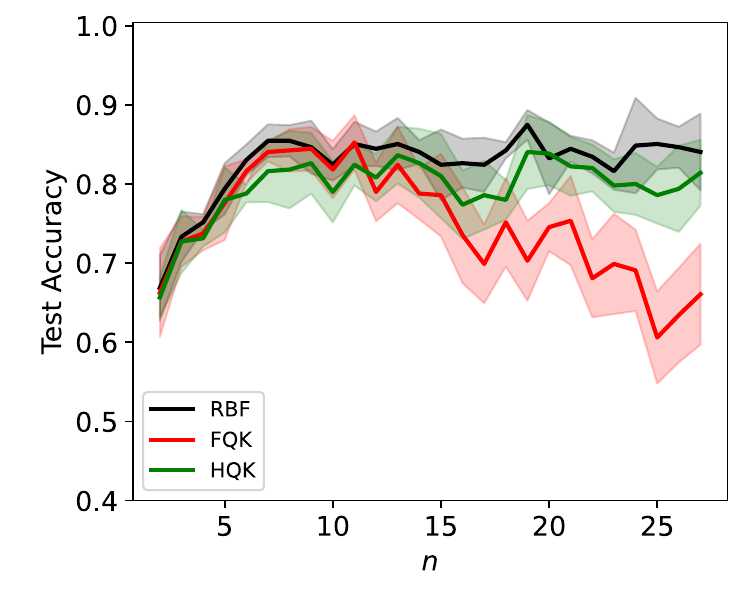}
        \put(-3,75){\scriptsize(b)}
    \end{overpic}

    \caption{Same as Figs.~\ref{fig:plots01} and \ref{fig:plots06},
    for classification of odd and even MNIST digits. (a) 
    The optimal value of 
    the HQK hyperparameter
    $\lambda$ selected for this dataset is $0.30$ (vertical dashed line). (b) 
    Test accuracy peaks at approximately $0.85$ for the RBF kernel. This is due to the limited number of training samples, with an average of only $10$ samples per digit within each class, making classification more challenging. However, the comparative behavior of FQK and HQK remains consistent: FQK accuracy drops for larger $n$, whereas HQK remains stable. Both HQK and RBF plateau at high $n$, while FQK shows a continued downward trend.}
    \label{fig:plotsoe}
\end{figure}

We replicate the previous analysis by performing a hyperparameter sweep over the same qubit counts to identify the range of $\lambda$ that maximizes cross-validation accuracy (see Fig.~\ref{fig:plotsoe-a}). For 
numbers of qubits
larger than $12$, the 
results indicate a pronounced maximum of
the cross-validation accuracy and hence a superior performance of HQK with respect to FQK (which, remember, corresponds to the limit $\lambda\to\infty$ of HQK).  
Values of $\lambda$ within the range $[0.05, 0.3]$ 
consistently achieve high accuracy. 
The optimal value of $\lambda$ selected for this dataset is $0.3$.

    

A comparative analysis of accuracy on the unseen test data is shown in Fig.~\ref{fig:plotsoe-b}. The overall accuracy is lower compared to the $0/1$ and $0/6$ datasets for both quantum and classical kernels. This is due to the nature of the dataset: it consists of approximately $50$ training samples per class, with an average of only $10$ samples per digit within each class. As a result, the test accuracy is lower even for the classical RBF kernel.

Moreover, FQK maintains accuracy comparable to HQK only up to $13$ qubits, beyond which FQK exhibits a consistent decrease in accuracy with increasing qubit count just like in the $0/6$ simulation. The accuracy of HQK is, on average, $10\%$ higher than that of FQK for qubit counts greater than $20$. Notably, both RBF and HQK plateau at higher qubit counts, whereas FQK shows a continued downward trend. The behaviour of HQK across all subsets of the classical dataset underscores its robustness as a strategy for mitigating exponential concentration.

\subsection{Kernel Quality for Quantum Data}

A binary classification is performed on the dataset generated by the quantum circuits $V_0$ and $V_1$, see Fig.~\ref{fig:qdata}. The dataset consists of 100 training samples and 100 test samples.
For both sets, 50 samples are drawn from circuit $V_0$ and 50 from circuit $V_1$.

    
    

\begin{figure}[htbp]
    \centering

    \refstepcounter{mysubfig}\label{fig:plotszz-a}%
    \begin{overpic}[width=\linewidth]{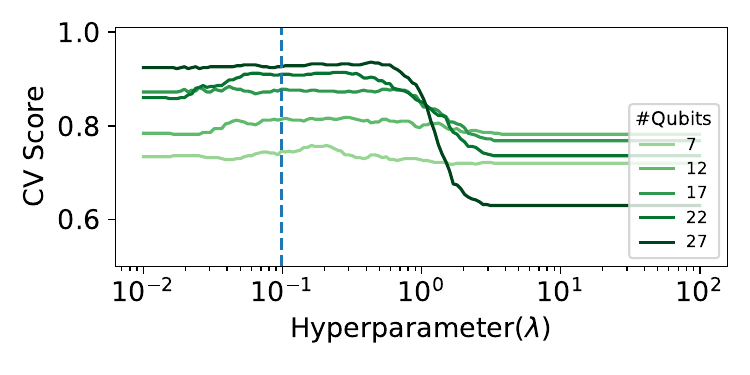}
        \put(-4,45){\scriptsize(a)}
    \end{overpic}

    \vspace{0.5em}

    \refstepcounter{mysubfig}\label{fig:plotszz-b}%
    \begin{overpic}[width=\linewidth]{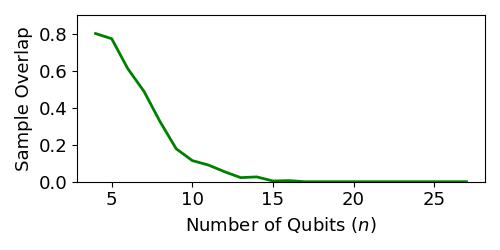}
        \put(-4,45){\scriptsize(b)}
    \end{overpic}

    \vspace{0.5em}

    \refstepcounter{mysubfig}\label{fig:plotszz-c}%
    \begin{overpic}[width=\linewidth]{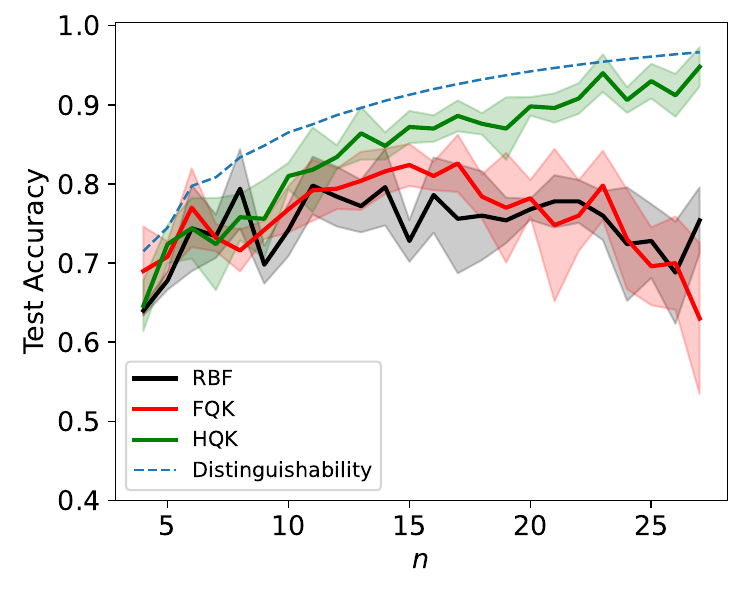}
        \put(-4,75){\scriptsize(c)}
    \end{overpic}

    \caption{Comparison of test scores 
    for classification of quantum circuit-generated data. (a) HQK hyperparameter optimization: Cross-validation scores 
    as a function of $\lambda$ for different numbers of qubits (see inset). The optimal hyperparameter selected is $\lambda = 0.1$ (vertical dashed line). (b) Fraction of data points shared between the training and the test set, both consisting of 100 randomly drawn samples. It decreases as the Hilbert space grows with increasing number $n$ of qubits. (c) Test accuracy for all kernels as a function of number of qubits $n$, alongside the distinguishability (theoretical upper bound for class distinction). The RBF kernel yields the lowest performance for up to $26$ qubits, while the HQK outperforms the RBF and FQK at large $n$, demonstrating effective mitigation of exponential concentration and approaching the theoretical accuracy limit. Notably, the HQK not only surpasses both the FQK and RBF kernel in test accuracy, but also achieves performance close to the theoretical maximum.}
    \label{fig:plotszz}
\end{figure}

A hyperparameter sweep is performed over $\lambda$ to identify the value that maximizes the cross-validation accuracy. In Fig.~\ref{fig:plotszz-a}, we see that  
the highest CV scores are reached
in the range $\lambda \in [0.01, 0.5]$. Across all qubit counts, the accuracy in this regime consistently exceeds that of the fidelity-based quantum kernel
(corresponding to the limit $\lambda\to\infty)$,
with the performance gap widening as the number of qubits increases. Based on 
these results, the optimal hyperparameter for this dataset is selected as $\lambda = 0.1$. Notably, unlike the behaviour observed in the MNIST data, the cross-validation score remains stable 
in the regime of small $\lambda$.
As we have checked, 
this stability persists 
down to $\lambda = 10^{-15}$, at which point computational precision becomes the limiting factor.

A robust assessment of model quality requires that the test samples be distinct from those used during training. To quantify this, we compute the overlap between the training and test sets across varying numbers of qubits. As shown in Fig.~\ref{fig:plotszz-b}, the fraction of samples shared between the two sets is $80\%$ at $4$ qubits but decreases as the number of qubits increases, reaching $0\%$ for systems exceeding $15$ qubits. This ensures that, at sufficiently large qubit counts, the model cannot rely on memorization and must instead generalize the underlying structure of the problem.

Finally, we perform a scaling analysis on the dataset as we increase the number of qubits from $4$ to $27$ (see Fig.~\ref{fig:plotszz-c}). The plot compares the test accuracy of the fidelity quantum kernel (FQK), the Hamming quantum kernel (HQK), and the classical Gaussian RBF kernel. We also include the distinguishability curve, which shows the maximum possible accuracy for this dataset. 
The distinguishability is defined as
\begin{equation}
D(P_0, P_1)= \sum_{b} \frac{1}{2} \max(P_0(b), P_1(b))
\end{equation}
Here, $P_0$ and $P_1$ denote the probabilities of sampling a bitstring $b$ from circuits $V_0$ and $V_1$, respectively, as computed via statevector simulation. Distinguishability represents the expected classification accuracy 
of an optimal classifier, assuming the underlying probability distributions are known and no prior bias is present. 
The optimal classifier, which reaches the best possible accuracy, is the one that assigns bitstring $b$ to $V_0$ if $P_0(b) > P_1(b)$, and to $V_1$ otherwise.
The distinguishability score takes values in the interval $\left[\tfrac{1}{2}, 1\right]$, where a score of $1$ indicates that the two distributions are perfectly distinguishable, while a score of $\tfrac{1}{2}$ implies that the distributions are statistically 
indistinguishable, i.e., identical.

Distinguishability is similar to another metric called the 
overlap coefficient \cite{inman89}.
\begin{equation}
OVL(P_0, P_1)= \sum_{b} \min(P_0(b), P_1(b)) ~.
\end{equation}
OVL is a normalized similarity measure that ranges from $0$ (no overlap) to $1$ (perfect overlap). It can be related to $D$ as follows: $D=1-OVL/2$.

The results agree with our expectation that quantum data is well suited to quantum kernels, as seen in the good performance of the FQK. The RBF kernel, however, performs similarly to the FQK and even slightly better for more than $25$ qubits. The HQK shows the strongest performance overall: it consistently beats both the FQK and the RBF kernel for all qubit counts, and the performance gap grows as the number of qubits increases. In addition, the HQK accuracy stays close to the theoretical maximum given by the distinguishability. This shows that the HQK can not only avoid exponential concentration, but at the same time it can beat well established classical models in certain tasks.


\section{Conclusion and Open Questions}
\label{conc}

Exponential concentration poses one of the most significant challenges to the large-scale deployment of quantum kernel methods for support vector machines. To address this issue, we propose the Hamming Quantum Kernel (HQK), a method that exploits the complete measurement statistics generated during the calculation of a kernel entry by computing a weighted sum of the probability amplitudes of output bitstrings, 
grouped according to their Hamming weight. The effectiveness and robustness of the proposed method are evaluated on a diverse set of classical and quantum datasets.

The Hamming quantum kernel (HQK) provides a clear advantage in test accuracy over the standard fidelity quantum kernel (FQK) for qubit counts greater than $15$ on both classical and quantum datasets. Our empirical results indicate that the HQK is a strong candidate to replace the FQK in large-scale settings or when computational resources are limited.

While the HQK achieves competitive accuracy on the MNIST dataset, approaching the performance of the classical RBF kernel with only a small gap for the $0/6$ and $0/1$ subsets, it demonstrates a decisive advantage on quantum data. For the quantum dataset, the FQK performs comparably to, and sometimes better than, the RBF kernel, highlighting how the nature of the dataset influences the effectiveness of (quantum) support vector machines. In this regime, the HQK strictly outperforms both the RBF and the FQK across all qubit counts, with the advantage growing as the number of qubits increases. Furthermore, the test accuracy of the HQK remains close to the maximum achievable accuracy.

No kernel alignment is performed for any dataset, making the HQK extraction process both problem-agnostic and efficient. 
Furthermore, the optimal hyperparameter value consistently peaks at approximately $\lambda = 0.1$ across all tested datasets, allowing the algorithm to be deployed directly without requiring a cross-validation step.

These promising results open several avenues for future investigation. The first and most fundamental direction is the development of a general 
framework for finding efficient distribution-informed classical post-processing functions, of which HQK represents a promising instance. 
Such a framework could systematically guide the design of scalable quantum kernels beyond the specific construction proposed here.
A second direction concerns the MNIST dataset: in particular, understanding the difference in accuracy achieved by the HQK between the odd/even task and the $0/6$ task, and whether targeted improvements can close this gap further. A third exciting direction stems from the strong performance of the HQK on quantum datasets, where understanding the factors that govern its performance across diverse quantum circuits could unlock further gains and broaden its applicability.

\begin{acknowledgments}
This research was carried out under the AI4QT project funded by EUREKA. We also thank Marco Dall'ara, Martin Koppenhöfer and Walter Hahn for many helpful discussions on this topic.
\end{acknowledgments}

\bibliography{citations}

@article{PhysRevLett.113.130503,
  title = {Quantum Support Vector Machine for Big Data Classification},
  author = {Rebentrost, Patrick and Mohseni, Masoud and Lloyd, Seth},
  journal = {Phys. Rev. Lett.},
  volume = {113},
  issue = {13},
  pages = {130503},
  numpages = {5},
  year = {2014},
  month = {Sep},
  publisher = {American Physical Society},
  doi = {10.1103/PhysRevLett.113.130503},
  url = {https://link.aps.org/doi/10.1103/PhysRevLett.113.130503}     
}

@article{Havl_ek_2019,
   title={Supervised learning with quantum-enhanced feature spaces},
   volume={567},
   ISSN={1476-4687},
   url={http://dx.doi.org/10.1038/s41586-019-0980-2}     ,
   DOI={10.1038/s41586-019-0980-2},
   number={7747},
   journal={Nature},
   publisher={Springer Science and Business Media LLC},
   author={Havlíček, Vojtěch and Córcoles, Antonio D. and Temme, Kristan and Harrow, Aram W. and Kandala, Abhinav and Chow, Jerry M. and Gambetta, Jay M.},
   year={2019},
   month=mar, pages={209–212} }

@article{bcancer,
author = { Saini, Shivani and
Khosla, PK and
Kaur, Manjit and
Singh, Gurmohan 
},
year = {2020},
month = {05},
title = { Quantum Driven Machine Learning},
volume = {59},
journal = {International Journal of Theoretical Physics
}
}

@article{Schnabel_2025,
   title={Quantum kernel methods under scrutiny: a benchmarking study},
   volume={7},
   ISSN={2524-4914},
   url={http://dx.doi.org/10.1007/s42484-025-00273-5}     ,
   DOI={10.1007/s42484-025-00273-5},
   number={1},
   journal={Quantum Machine Intelligence},
   publisher={Springer Science and Business Media LLC},
   author={Schnabel, Jan and Roth, Marco},
   year={2025},
   month=apr }

@article{Liu_2021,
   title={A rigorous and robust quantum speed-up in supervised machine learning},
   volume={17},
   ISSN={1745-2481},
   url={http://dx.doi.org/10.1038/s41567-021-01287-z}            ,
   DOI={10.1038/s41567-021-01287-z},
   number={9},
   journal={Nature Physics},
   publisher={Springer Science and Business Media LLC},
   author={Liu, Yunchao and Arunachalam, Srinivasan and Temme, Kristan},
   year={2021},
   month=jul, pages={1013–1017} }

@misc{thanasilp2024exponentialconcentrationquantumkernel,
      title={Exponential concentration in quantum kernel methods}, 
      author={Supanut Thanasilp and Samson Wang and M. Cerezo and Zoë Holmes},
      year={2024},
      eprint={2208.11060},
      archivePrefix={arXiv},
      primaryClass={quant-ph},
      url={https://arxiv.org/abs/2208.11060}, 
}

@misc{muser2023provableadvantageskernelbasedquantum,
      title={Provable advantages of kernel-based quantum learners and quantum preprocessing based on Grover's algorithm}, 
      author={Till Muser and Elias Zapusek and Vasilis Belis and Florentin Reiter},
      year={2023},
      eprint={2309.14406},
      archivePrefix={arXiv},
      primaryClass={quant-ph},
      url={https://arxiv.org/abs/2309.14406}, 
}

@article{Huang_2021,
   title={Power of data in quantum machine learning},
   volume={12},
   ISSN={2041-1723},
   url={http://dx.doi.org/10.1038/s41467-021-22539-9}          ,
   DOI={10.1038/s41467-021-22539-9},
   number={1},
   journal={Nature Communications},
   publisher={Springer Science and Business Media LLC},
   author={Huang, Hsin-Yuan and Broughton, Michael and Mohseni, Masoud and Babbush, Ryan and Boixo, Sergio and Neven, Hartmut and McClean, Jarrod R.},
   year={2021},
   month=may }

@article{Slattery_2023,
   title={Numerical evidence against advantage with quantum fidelity kernels on classical data},
   volume={107},
   ISSN={2469-9934},
   url={http://dx.doi.org/10.1103/PhysRevA.107.062417}          ,
   DOI={10.1103/physreva.107.062417},
   number={6},
   journal={Physical Review A},
   publisher={American Physical Society (APS)},
   author={Slattery, Lucas and Shaydulin, Ruslan and Chakrabarti, Shouvanik and Pistoia, Marco and Khairy, Sami and Wild, Stefan M.},
   year={2023},
   month=jun }

@misc{agliardi2024mitigatingexponentialconcentrationcovariant,
      title={Mitigating exponential concentration in covariant quantum kernels for subspace and real-world data}, 
      author={Gabriele Agliardi and Giorgio Cortiana and Anton Dekusar and Kumar Ghosh and Naeimeh Mohseni and Corey O'Meara and Víctor Valls and Kavitha Yogaraj and Sergiy Zhuk},
      year={2024},
      eprint={2412.07915},
      archivePrefix={arXiv},
      primaryClass={quant-ph},
      url={https://arxiv.org/abs/2412.07915}, 
}

@article{Bharti_2022,
   title={Noisy intermediate-scale quantum algorithms},
   volume={94},
   ISSN={1539-0756},
   url={http://dx.doi.org/10.1103/RevModPhys.94.015004}       ,
   DOI={10.1103/revmodphys.94.015004},
   number={1},
   journal={Reviews of Modern Physics},
   publisher={American Physical Society (APS)},
   author={Bharti, Kishor and Cervera-Lierta, Alba and Kyaw, Thi Ha and Haug, Tobias and Alperin-Lea, Sumner and Anand, Abhinav and Degroote, Matthias and Heimonen, Hermanni and Kottmann, Jakob S. and Menke, Tim and Mok, Wai-Keong and Sim, Sukin and Kwek, Leong-Chuan and Aspuru-Guzik, Alán},
   year={2022},
   month=feb }

@article{Hofmann_2008,
   title={Kernel methods in machine learning},
   volume={36},
   ISSN={0090-5364},
   url={http://dx.doi.org/10.1214/009053607000000677}      ,
   DOI={10.1214/009053607000000677},
   number={3},
   journal={The Annals of Statistics},
   publisher={Institute of Mathematical Statistics},
   author={Hofmann, Thomas and Schölkopf, Bernhard and Smola, Alexander J.},
   year={2008},
   month=jun }

@InProceedings{10.1007/11776420_14, 
author="Minh, Ha Quang
and Niyogi, Partha
and Yao, Yuan",
editor="Lugosi, G{\'a}bor
and Simon, Hans Ulrich",
title="Mercer's Theorem, Feature Maps, and Smoothing",
booktitle="Learning Theory",
year="2006",
publisher="Springer Berlin Heidelberg",
address="Berlin, Heidelberg",
pages="154--168",
isbn="978-3-540-35296-9"
}

@inproceedings{NIPS2001_1f71e393,
 author = {Cristianini, Nello and Shawe-Taylor, John and Elisseeff, Andr\'{e} and Kandola, Jaz},
 booktitle = {Advances in Neural Information Processing Systems},
 editor = {T. Dietterich and S. Becker and Z. Ghahramani},
 pages = {},
 publisher = {MIT Press},
 title = {On Kernel-Target Alignment},
 url = {https://proceedings.neurips.cc/paper_files/paper/2001/file/1f71e393b3809197ed66df836fe833e5-Paper.pdf},
 volume = {14},
 year = {2001}
}

@book{Rasmussen2006Gaussian,
  author = {Rasmussen, Carl Edward and Williams, Christopher K. I.},
  publisher = {The MIT Press},
  title = {Gaussian Processes for Machine Learning},
  year = 2006
}

@ARTICLE{6296535,
  author={Deng, Li},
  journal={IEEE Signal Processing Magazine}, 
  title={The MNIST Database of Handwritten Digit Images for Machine Learning Research [Best of the Web]}, 
  year={2012},
  volume={29},
  number={6},
  pages={141-142},
  doi={10.1109/MSP.2012.2211477}}

@article{kfold,
  author = {Victor Wandera Lumumba and Dennis Kiprotich and Mary Lemasulani Mpaine and Njoka Grace Makena and Musyimi Daniel Kavita},
  title = {Comparative Analysis of Cross-Validation Techniques: LOOCV, K-folds Cross-Validation, and Repeated K-folds Cross-Validation in Machine Learning Models
},
  journal = {American Journal of Theoretical and Applied Statistics},
  volume = {13},
  number = {5},
  pages = {127-137},
  doi = {10.11648/j.ajtas.20241305.13},
  url = {https://doi.org/10.11648/j.ajtas.20241305.13},
 year = {2024}
}

@article{MACKIEWICZ1993303,
title = {Principal components analysis (PCA)},
journal = {Computers \& Geosciences},
volume = {19},
number = {3},
pages = {303-342},
year = {1993},
issn = {0098-3004},
doi = {https://doi.org/10.1016/0098-3004(93)90090-R},
url = {https://www.sciencedirect.com/science/article/pii/009830049390090R},
author = {Andrzej Maćkiewicz and Waldemar Ratajczak}
}

@misc{barenco1996stabilisationquantumcomputationssymmetrisation,
      title={Stabilisation of Quantum Computations by Symmetrisation}, 
      author={Adriano Barenco and Andre` Berthiaume and David Deutsch and Artur Ekert and Richard Jozsa and Chiara Macchiavello},
      year={1996},
      eprint={quant-ph/9604028},
      archivePrefix={arXiv},
      primaryClass={quant-ph},
      url={https://arxiv.org/abs/quant-ph/9604028}, 
}

@Article{Larocca2025,
author={Larocca, Mart{\'i}n
and Thanasilp, Supanut
and Wang, Samson
and Sharma, Kunal
and Biamonte, Jacob
and Coles, Patrick J.
and Cincio, Lukasz
and McClean, Jarrod R.
and Holmes, Zo{\"e}
and Cerezo, M.},
title={Barren plateaus in variational quantum computing},
journal={Nature Reviews Physics},
year={2025},
month={Apr},
day={01},
volume={7},
number={4},
pages={174-189},
issn={2522-5820},
doi={10.1038/s42254-025-00813-9},
url={https://doi.org/10.1038/s42254-025-00813-9}  
}

@article{Belis_2024,
   title={Quantum anomaly detection in the latent space of proton collision events at the LHC},
   volume={7},
   ISSN={2399-3650},
   url={http://dx.doi.org/10.1038/s42005-024-01811-6}    ,
   DOI={10.1038/s42005-024-01811-6},
   number={1},
   journal={Communications Physics},
   publisher={Springer Science and Business Media LLC},
   author={Belis, Vasilis and Woźniak, Kinga Anna and Puljak, Ema and Barkoutsos, Panagiotis and Dissertori, Günther and Grossi, Michele and Pierini, Maurizio and Reiter, Florentin and Tavernelli, Ivano and Vallecorsa, Sofia},
   year={2024},
   month=Oct }

@article{inman89,
author = {Henry F. Inman and Edwin L. Bradley Jr},
title = {The overlapping coefficient as a measure of agreement between probability distributions and point estimation of the overlap of two normal densities},
journal = {Communications in Statistics - Theory and Methods},
volume = {18},
number = {10},
pages = {3851--3874},
year = {1989},
publisher = {Taylor \& Francis},
doi = {10.1080/03610928908830127},
URL = { 
        https://doi.org/10.1080/03610928908830127   
},
eprint = { 
        https://doi.org/10.1080/03610928908830127   
}
}

@misc{aminpour2024strategicdatareuploadspathway,
      title={Strategic Data Re-Uploads: A Pathway to Improved Quantum Classification Data Re-Uploading Strategies for Improved Quantum Classifier Performance}, 
      author={S. Aminpour and Y. Banad and S. Sharif},
      year={2024},
      eprint={2405.09377},
      archivePrefix={arXiv},
      primaryClass={quant-ph},
      url={https://arxiv.org/abs/2405.09377}, 
}

@article{schuld2019,
  title = {Quantum Machine Learning in Feature Hilbert Spaces},
  author = {Schuld, Maria and Killoran, Nathan},
  journal = {Phys. Rev. Lett.},
  volume = {122},
  issue = {4},
  pages = {040504},
  numpages = {6},
  year = {2019},
  month = {Feb},
  publisher = {American Physical Society},
  doi = {10.1103/PhysRevLett.122.040504},
  url = {https://link.aps.org/doi/10.1103/PhysRevLett.122.040504}   
}

\end{document}